%% file: main.tex
\documentclass{article}

\usepackage[final]{style/neurips_2022_ml4ps}

\usepackage[utf8]{inputenc} 
\usepackage[T1]{fontenc}    
\usepackage{hyperref}       
\usepackage{url}            
\usepackage{booktabs}       
\usepackage{amsfonts}       
\usepackage{nicefrac}       
\usepackage{microtype}      
\usepackage{xcolor}         
\usepackage{amsmath}
\usepackage{amssymb}
\usepackage{bm}
\usepackage{multirow}
\usepackage{adjustbox}
\usepackage{makecell}
\usepackage{lipsum}
\usepackage[para]{footmisc}
\usepackage{soul}           

\input{defs}

\title{Generating astronomical spectra from photometry with conditional diffusion models}

\author{%
  Lars Doorenbos \\
  University of Bern\\
  Bern, Switzerland \\
  \texttt{lars.doorenbos@unibe.ch} \\
   \And
  Stefano Cavuoti \\
  Astronomical Observatory of Capodimonte\\
  Naples, Italy \\
  \texttt{stefano.cavuoti@inaf.it} \\
   \And
  Giuseppe Longo \\
  University Federico II \\
  Naples, Italy \\
  \texttt{giuseppe.longo@unina.it}
   \And
  Massimo Brescia \\
  University Federico II\\
  Naples, Italy \\
  \texttt{massimo.brescia@unina.it}
   \And
  Raphael Sznitman \\
  University of Bern\\
  Bern, Switzerland \\
  \texttt{raphael.sznitman@unibe.ch} \\
   \And
  Pablo Márquez-Neila \\
  University of Bern\\
  Bern, Switzerland \\
  \texttt{pablo.marquez@unibe.ch} \\
}

\begin{document}

\maketitle

\begin{abstract}
A trade-off between speed and information controls our understanding of astronomical objects. Fast-to-acquire photometric observations provide global properties, while costly and time-consuming spectroscopic measurements enable a better understanding of the physics governing their evolution. Here, we tackle this problem by generating spectra directly from photometry, through which we obtain an estimate of their intricacies from easily acquired images. This is done by using multi-modal conditional diffusion models, where the best out of the generated spectra is selected with a contrastive network. Initial experiments on minimally processed SDSS galaxy data show promising results.
\end{abstract}

\section{Introduction}
Modern digital multi-band astronomical surveys are producing large volumes of photometric data, and will do so even more in the future. In most cases the scientific exploitation of the photometric data requires a vast amount of ancillary spectroscopic data, which can better characterize the sources. The latter, unfortunately, are much more demanding in terms of telescope time and can be acquired only for a small fraction of the photometric counterparts.
To quote an example, the Sloan Digital Sky Survey (SDSS, \cite{Ahumada2020}), which has one of the richest spectroscopic datasets to date, has observed more than one billion objects photometrically, while only about two million objects have been observed together with their spectra. 
The fact that multi-band photometry may be assimilated to low-resolution spectroscopy, motivates the search for methods capable of automatically guessing the shape of a spectrum, provided a significant number of examples. 
The possibility to use only photometric data to have a reliable guess of spectra would be useful for many applications, for instance to determine which objects are likely to be the most interesting, and need to be targeted for further investigations \cite{Masters2015}.

For our approach to this problem we make use of the paradigm of multi-modal learning. Recently, many multi-modal learning methods have combined all sorts of modalities \cite{bayoudh2022survey}, such as text-image \cite{radford2021learning}, text-video \cite{fan2019heterogeneous}, genome-image \cite{taleb2022contig}, and so on. In astronomy, \citet{wu2020predicting} attempted to predict galaxy spectra from images by predicting the latent variables of a variational autoencoder (VAE). A follow-up study \citet{holwerda2021predicting} aimed to confirm its practical use. 

In this work, we similarly attempt to generate galaxy spectra from images, by directly learning the relations between spectroscopic and photometric observations of the same object. In particular, we take inspiration from the text-image models CLIP~\cite{radford2021learning} and DALL-E~\cite{ramesh2021zero}. The images generated from text by the generative model DALL-E are ranked using the contrastive model CLIP to find the best fitting samples.\footnote{\url{https://openai.com/blog/dall-e}} We apply this concept to generate spectra from multi-band images, and find the best match with a contrastive network.\footnote{\url{https://github.com/LarsDoorenbos/generate-spectra}} 

\section{Method}

\paragraph{Data}
We extracted 64x64 thumbnails of the three inner SDSS bands (\textit{g}, \textit{r} and \textit{i}) for around 500 000 galaxies from the Sloan Digital Sky Survey Data Release 16 (SDSS DR16, \cite{Ahumada2020}), together with their corresponding spectra. As the SDSS pixel scale is 0.396 arcseconds per pixel, after random cropping, the images cover $\pm$22.85 arcseconds squared. Since the spectra covered different wavelength ($\lambda$) ranges, we discarded all spectra with a minimum $\log\lambda > 3.59$, or a maximum below 3.95. Then, we cropped every spectrum to the range between 3.59 and 3.95 and produced a uniform dataset of 3598-dimensional spectra. We normalized all spectra to the range $[-1, 1]$. In total, resulted in 496599 image-spectra pairs, which we split into a training set of 491599 and a validation set of 5000 pairs.

\subsection{Preliminaries}

\begin{figure}[t]
    \centering
    \includegraphics[width=\textwidth]{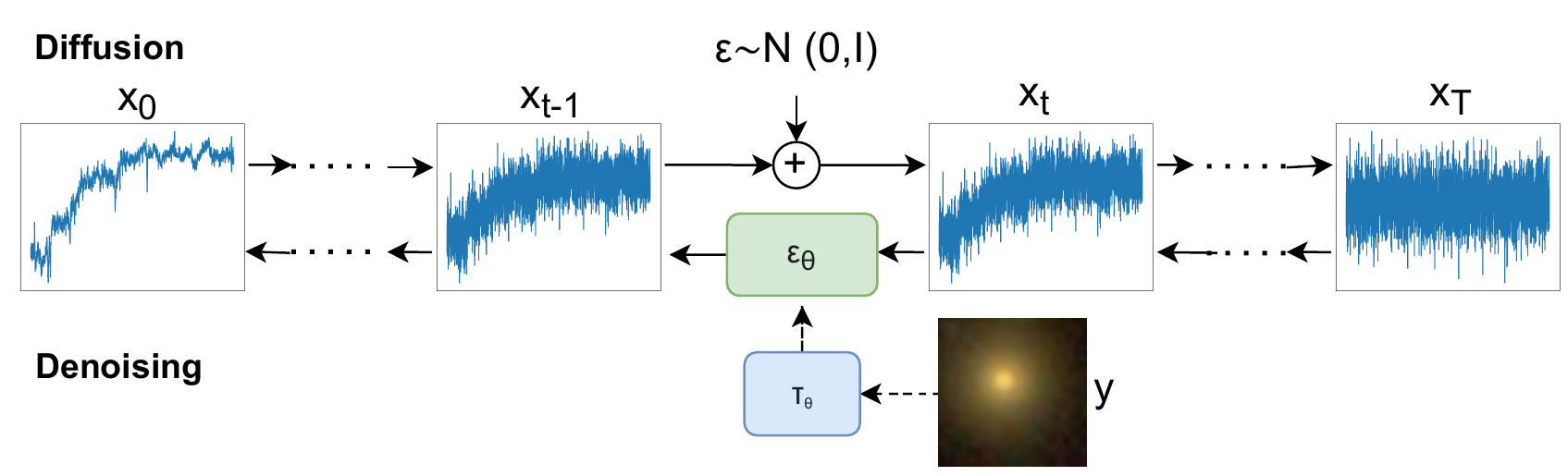}
    \caption{Training of the conditional diffusion model.}
    \label{fig:approach}
\end{figure}

\paragraph{Denoising diffusion probabilistic models (DDPM)} 

DDPMs, also known as diffusion models, are a class of generative models, that learn the data distribution through the combination of two parts: a diffusion and a denoising process \citet{ho2020denoising}. The diffusion process adds a certain amount of Gaussian noise to a sample, based on a set variance schedule $\beta_1,...,\beta_T$ at a timestep $t \in \left[ 1,T\right]$. This gradually transforms a sample $x_0$ into a standard Gaussian distribution at $x_T$.

The denoising process aims to reverse this process: given $t$, it predicts the denoised version $x_0$ from $x_t$ with an autoencoder neural network $\epsilon_\theta$. In practice, this is often learned by predicting the noise instead \citet{ho2020denoising,rombach2022high}, with the simplified objective
\begin{equation}
    L = \EX_{x_0,t,\epsilon\sim\mathcal{N}(0, \mathbf{I})} \left(\parallel \epsilon - \epsilon_\theta(x_t, t)\parallel^2 \right).
\end{equation}
From these processes, we can create new samples by sampling from a Gaussian distribution, then using the diffusion process to go back from $x_T$ to $x_0$.

\paragraph{Conditional diffusion models (CDM)}

While vanilla diffusion models are capable of modelling and sampling from a data distribution, we are interested in obtaining samples for a specific object. This can be achieved by sampling from a conditional distribution instead. This is implemented by adding a condition to the input of the noise prediction network. Given a condition $y$, we project it to a latent embedding using a learnable encoder $\tau_\theta$. 
This gives the loss function 
\begin{equation}
    L = \EX_{x_0,y,t,\epsilon\sim\mathcal{N}(0, \mathbf{I})} \left(\parallel \epsilon - \epsilon_\theta(x_t, t, \tau_\theta(y))\parallel^2 \right).
\end{equation}
We experimented with 2 forms of conditioning. The first projects $\tau_\theta(y)$ to the same dimensionality as $x_0$, and concatenates them channel-wise. The second follows \citet{rombach2022high}, and introduces the condition with cross-attention at multiple layers of the autoencoder.

\paragraph{Contrastive learning}

Contrastive learning is a framework for self-supervised learning. It optimizes a network by minimizing the latent distance between two views of the same object, while maximizing the distances to the latent codes of the other samples in the batch. This is done by minimizing the contrastive loss for mini-batches of size~$N$ \cite{chen2020simple},
\begin{equation}
\ell_{i, j} = -\log \frac{\exp (\text{sim}(\bm{z}_i, \bm{z}_j) / \tau)}{\sum^{2N}_{k=1} \bm{1}_{[k\neq i]} \exp (\text{sim}(\bm{z}_i, \bm{z}_k) / \tau)},
\end{equation}
for views $i$ and~$j$, where $\bm{z}$~represents their latent representation, $\tau$~the temperature, and $\text{sim}(\cdot,\cdot)$~the cosine similarity. In the present work, the two views of an object are its image and its spectrum.

\subsection{Generating spectra from images}
Combining the above, we train a CDM to learn the conditional distribution of spectra given an image. At inference time, we sample from this distribution a number of times in order to generate candidate spectra. 
Then, using the contrastive network, we compute the similarity between the latent representations of the generated spectra and that of the original image. The closest match is our best guess of the spectrum. 
An overview of the CDM training procedure is shown in Figure~\ref{fig:approach}.

\section{Experiments}
\label{sec:exp}

\begin{figure}[t]
  \centering
  \setlength\tabcolsep{0pt}
  \begin{tabular}{ccccc}
    \includegraphics[width=0.08\linewidth]{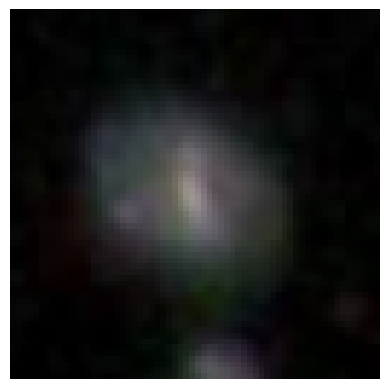} &
    \includegraphics[width=0.23\linewidth]{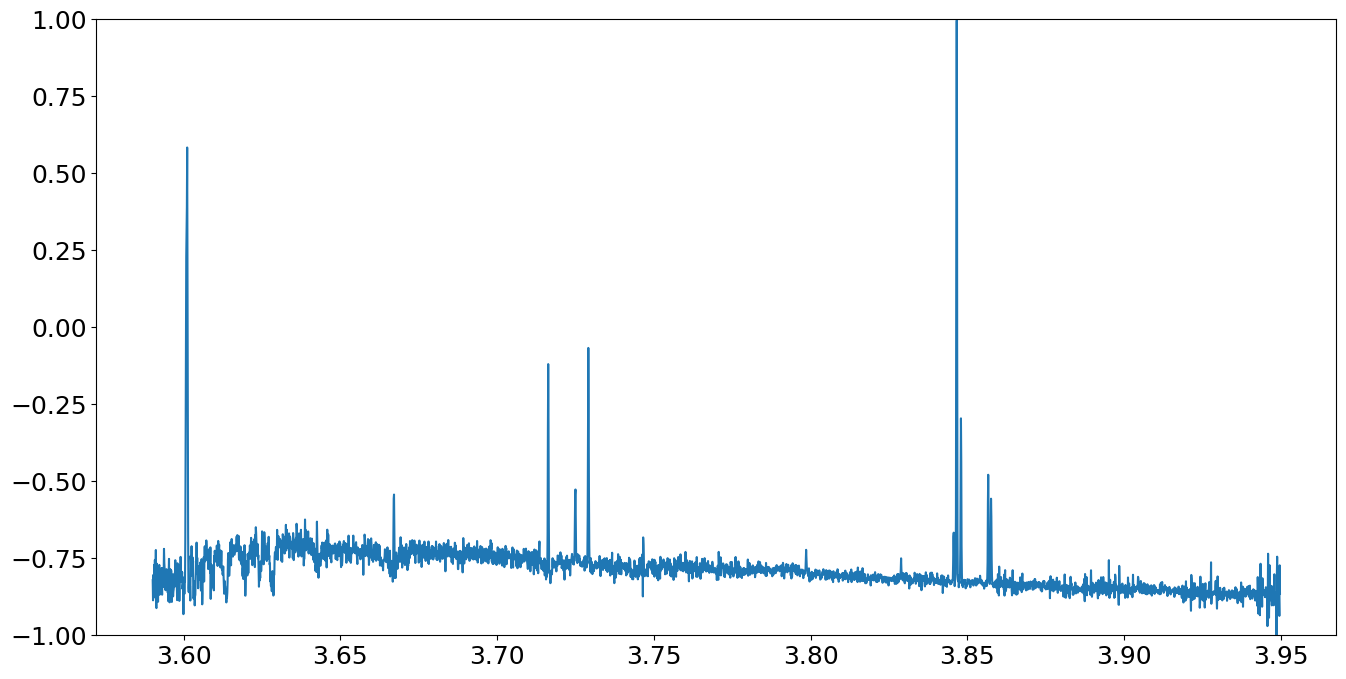} &
    \includegraphics[width=0.23\linewidth]{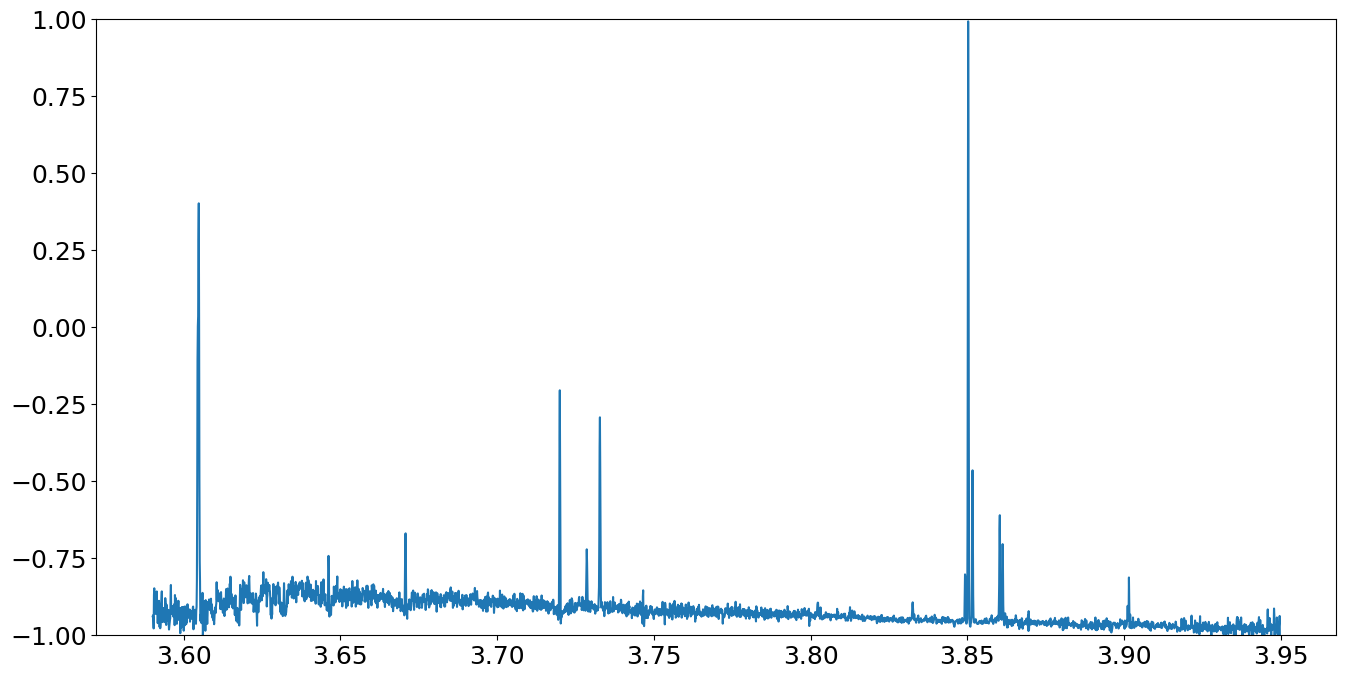} &
    \includegraphics[width=0.23\linewidth]{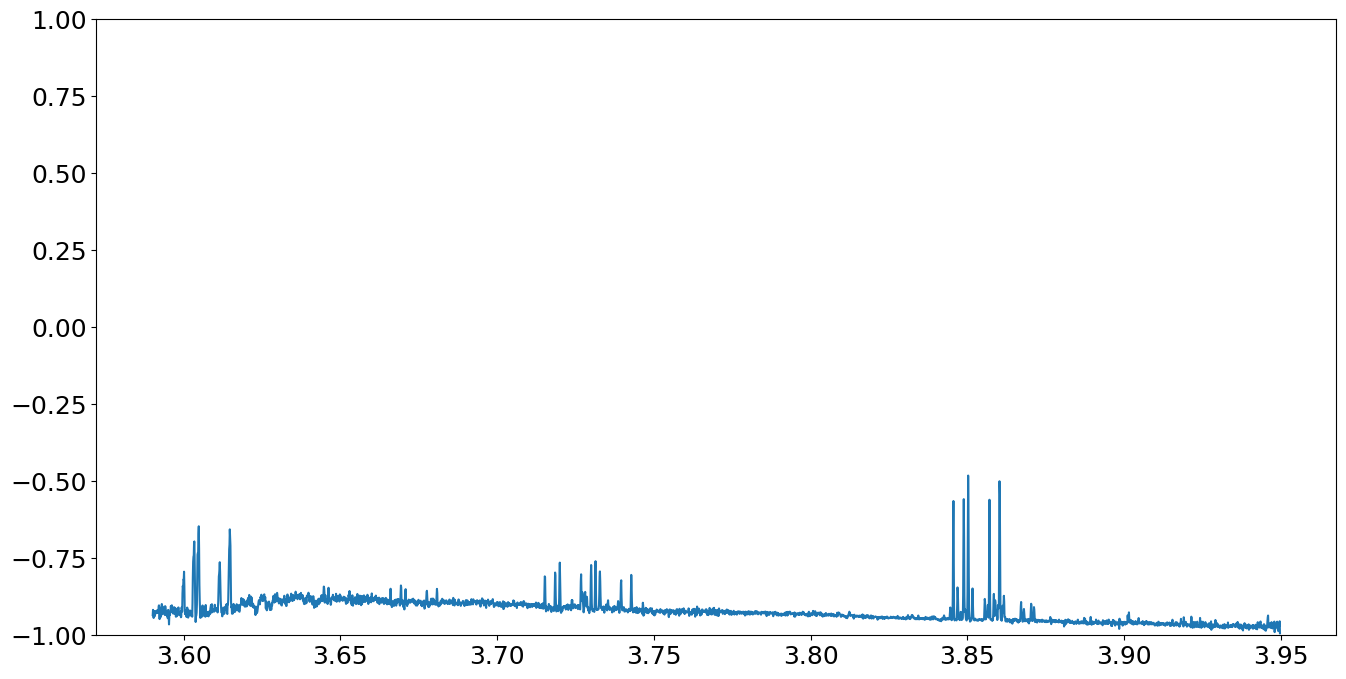} & 
    \includegraphics[width=0.23\linewidth]{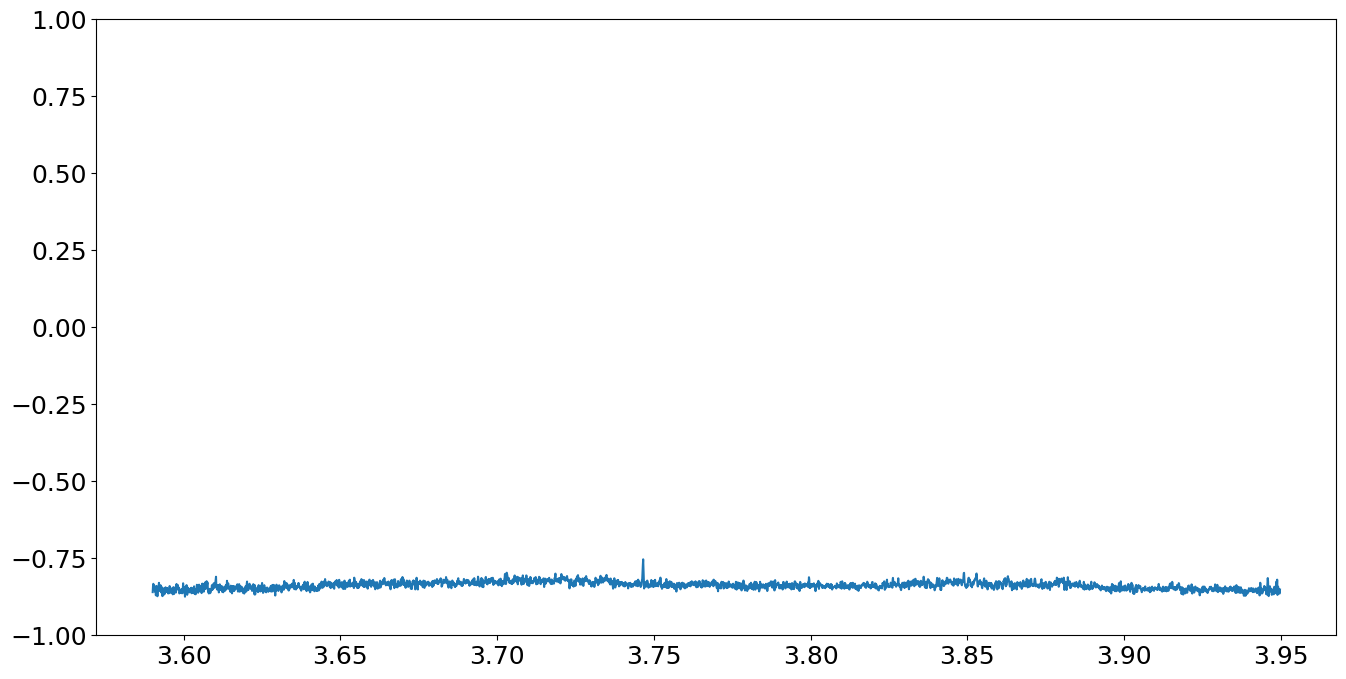} \\
    \includegraphics[width=0.08\linewidth]{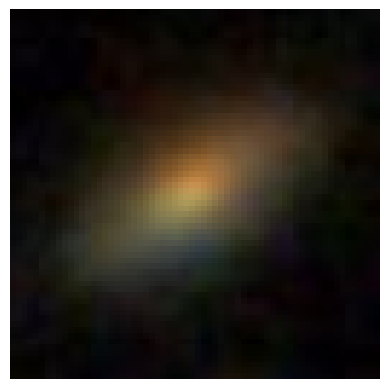} &
    \includegraphics[width=0.23\linewidth]{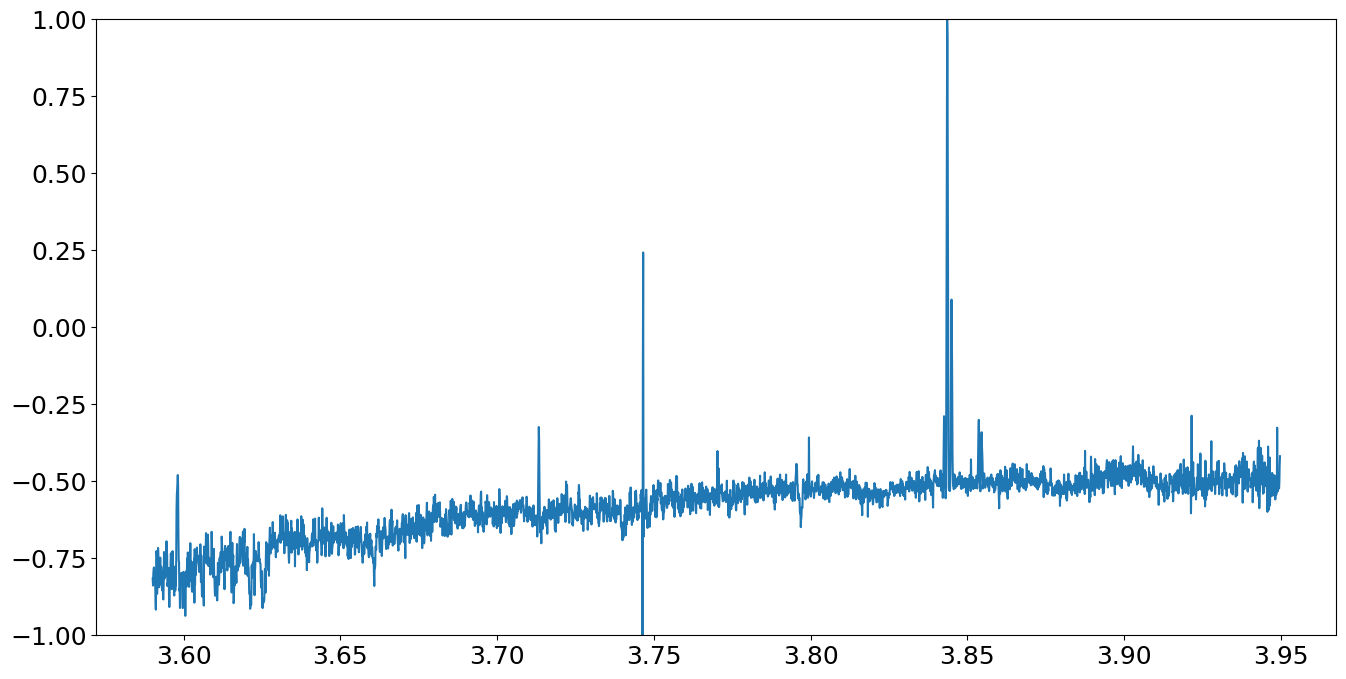} &
    \includegraphics[width=0.23\linewidth]{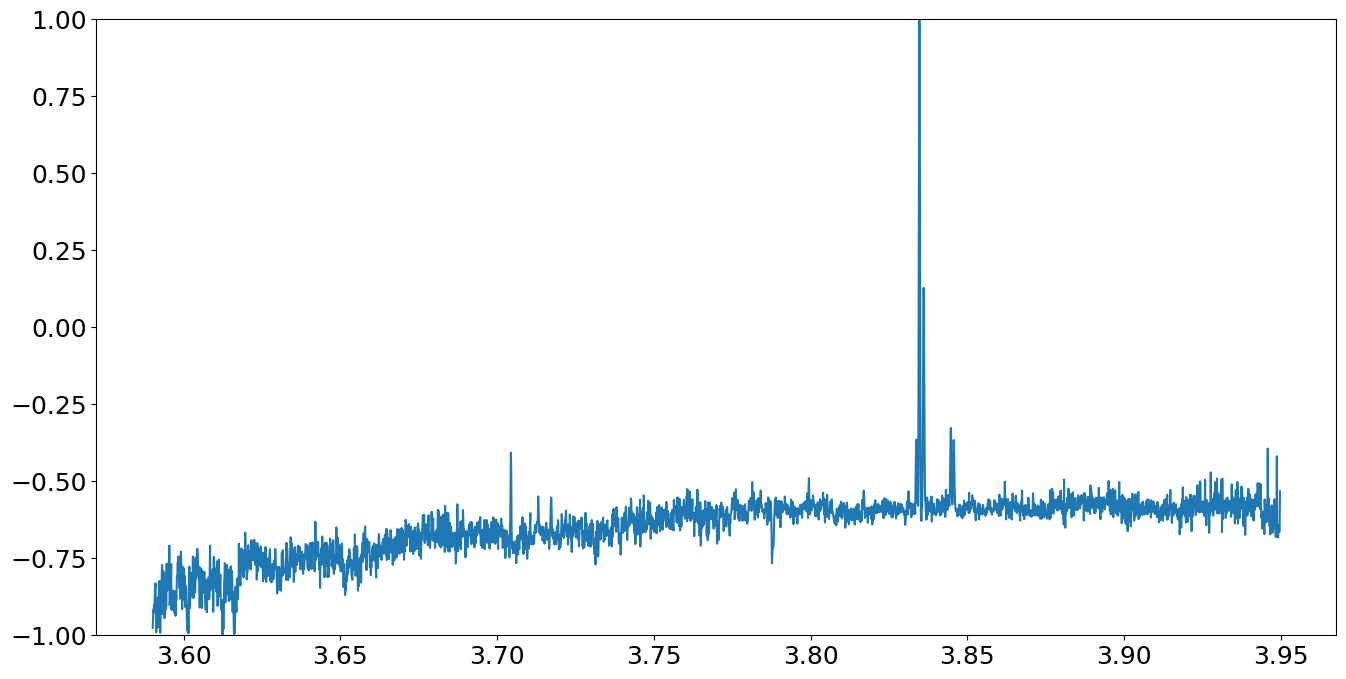} &
    \includegraphics[width=0.23\linewidth]{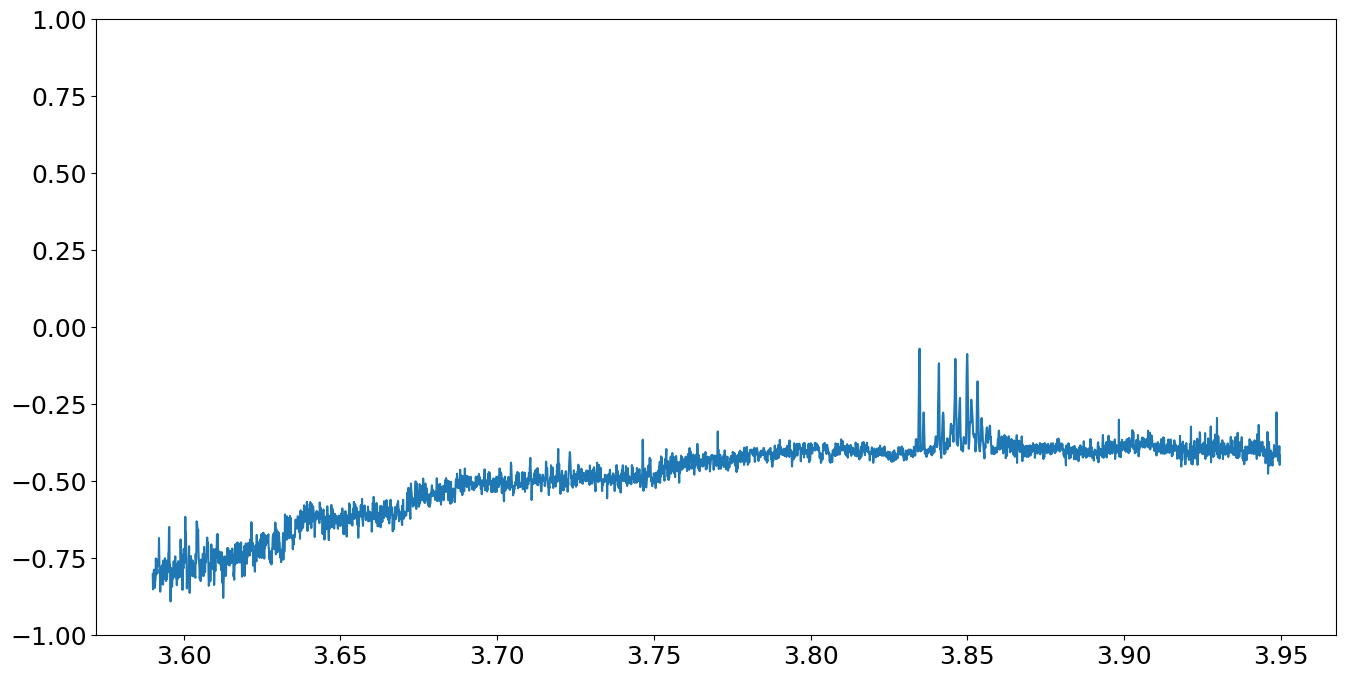} &
    \includegraphics[width=0.23\linewidth]{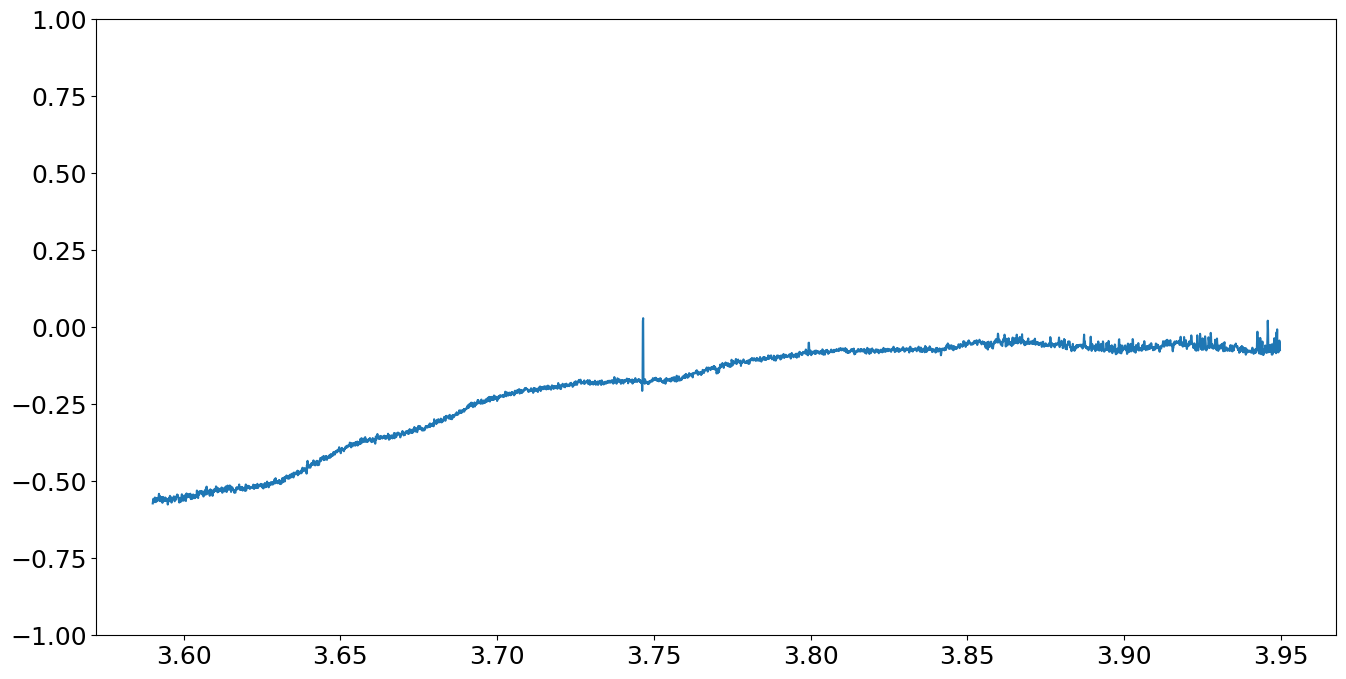} \\
    \includegraphics[width=0.08\linewidth]{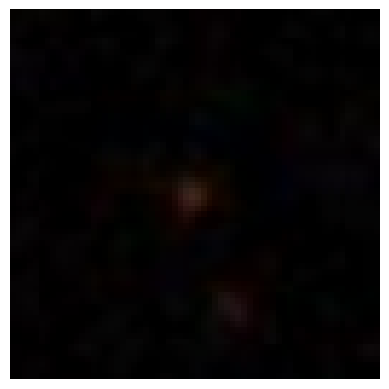} &
    \includegraphics[width=0.23\linewidth]{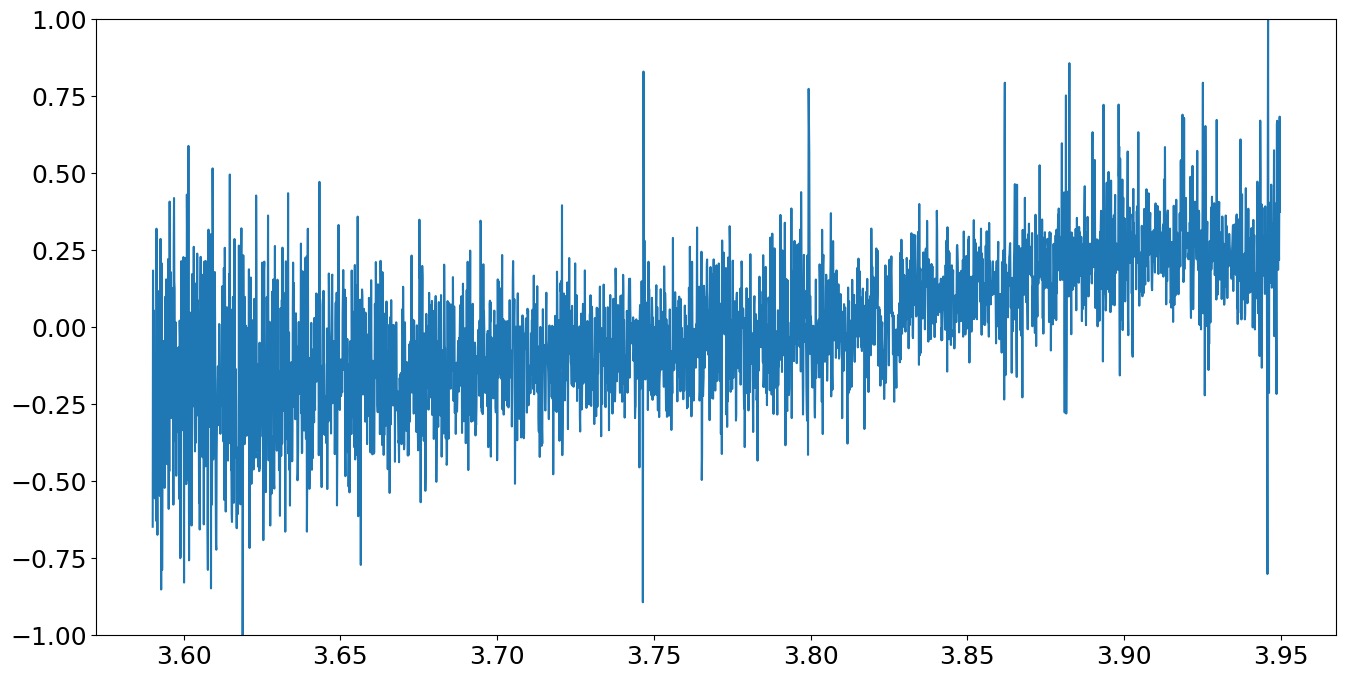} &
    \includegraphics[width=0.23\linewidth]{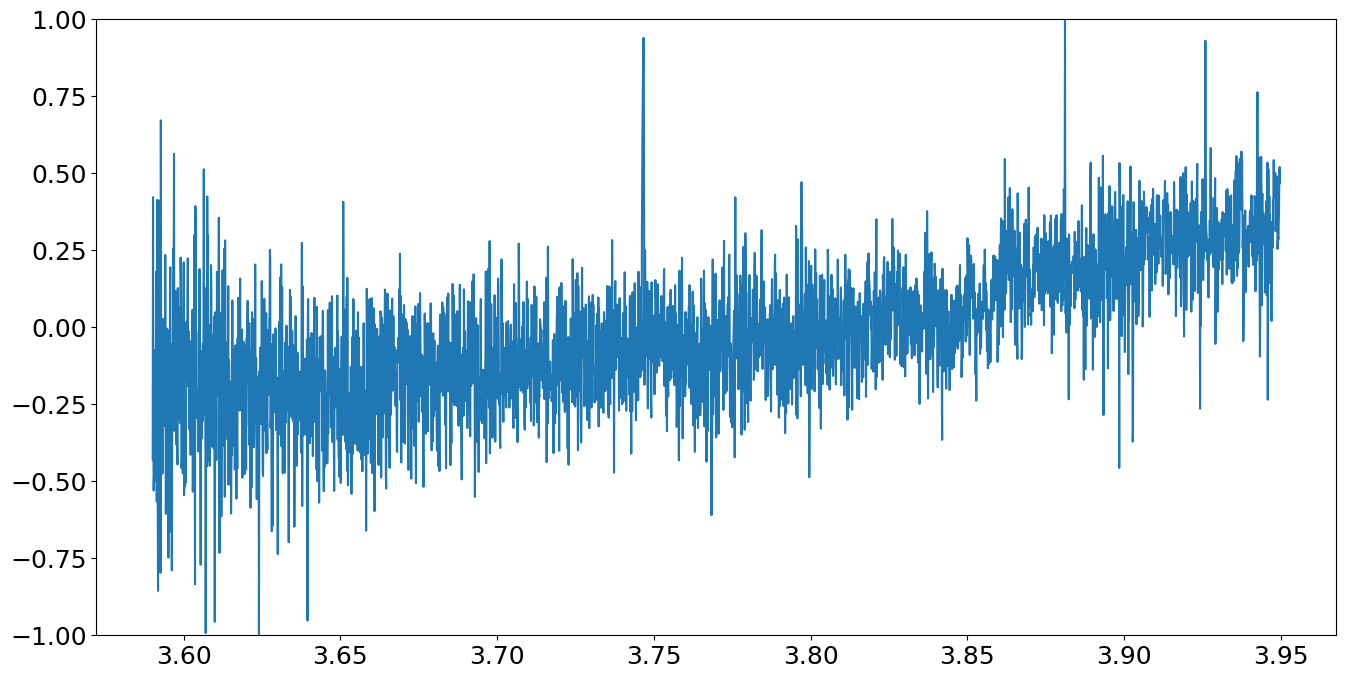} &
    \includegraphics[width=0.23\linewidth]{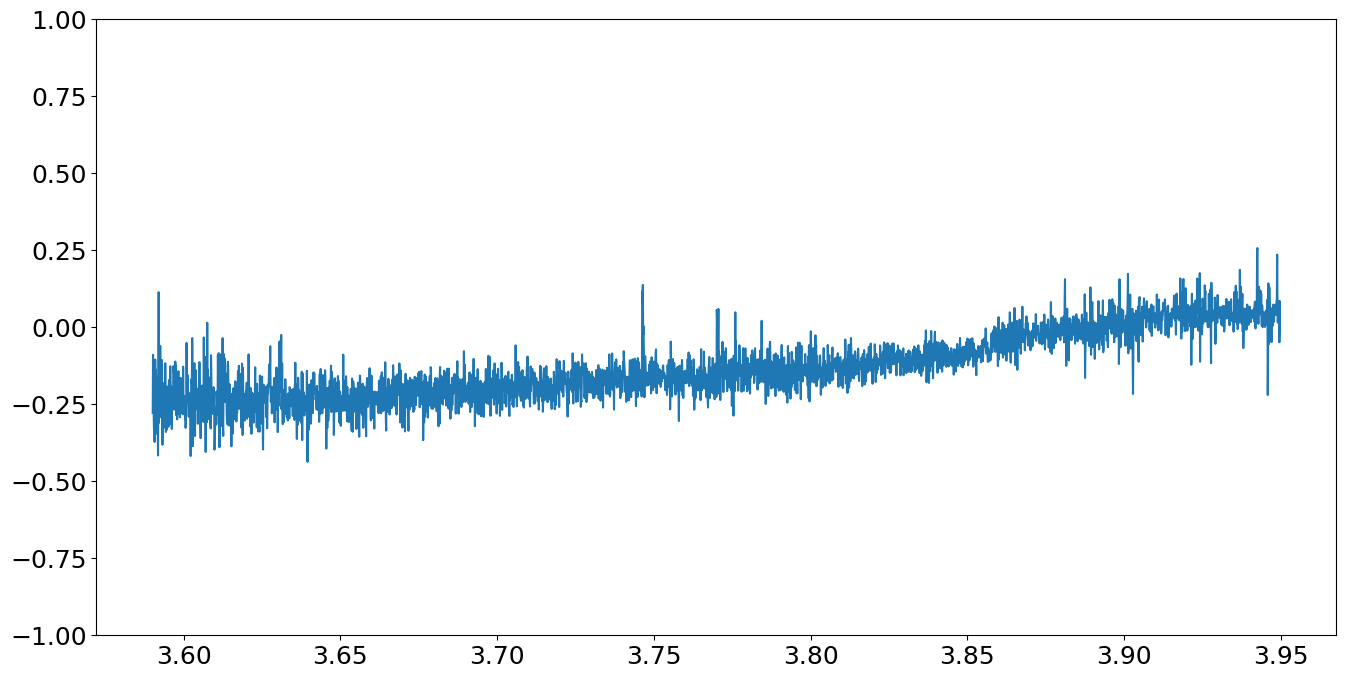}  &
    \includegraphics[width=0.23\linewidth]{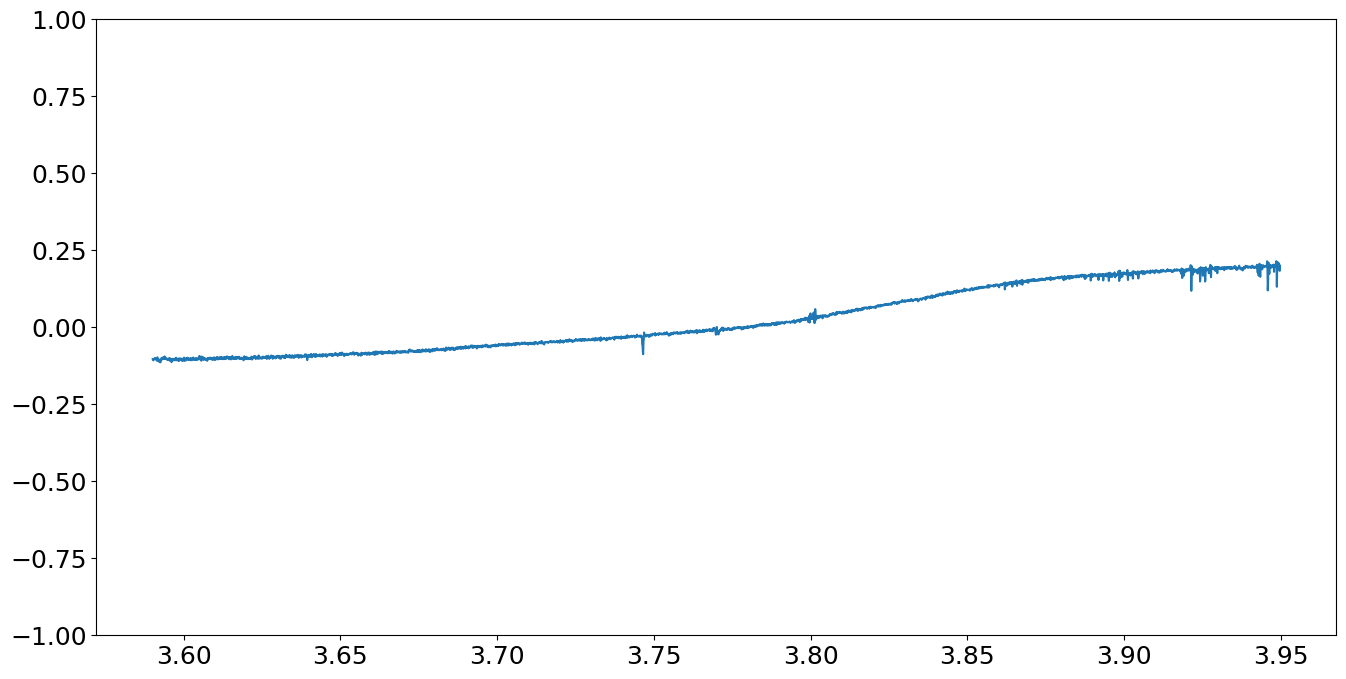} \\
    \includegraphics[width=0.08\linewidth]{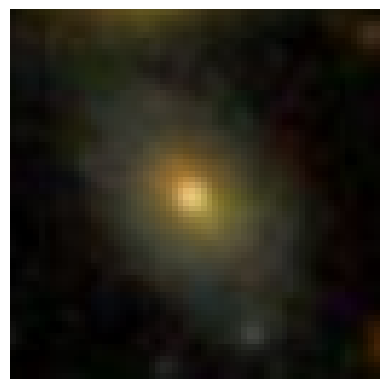} &
    \includegraphics[width=0.23\linewidth]{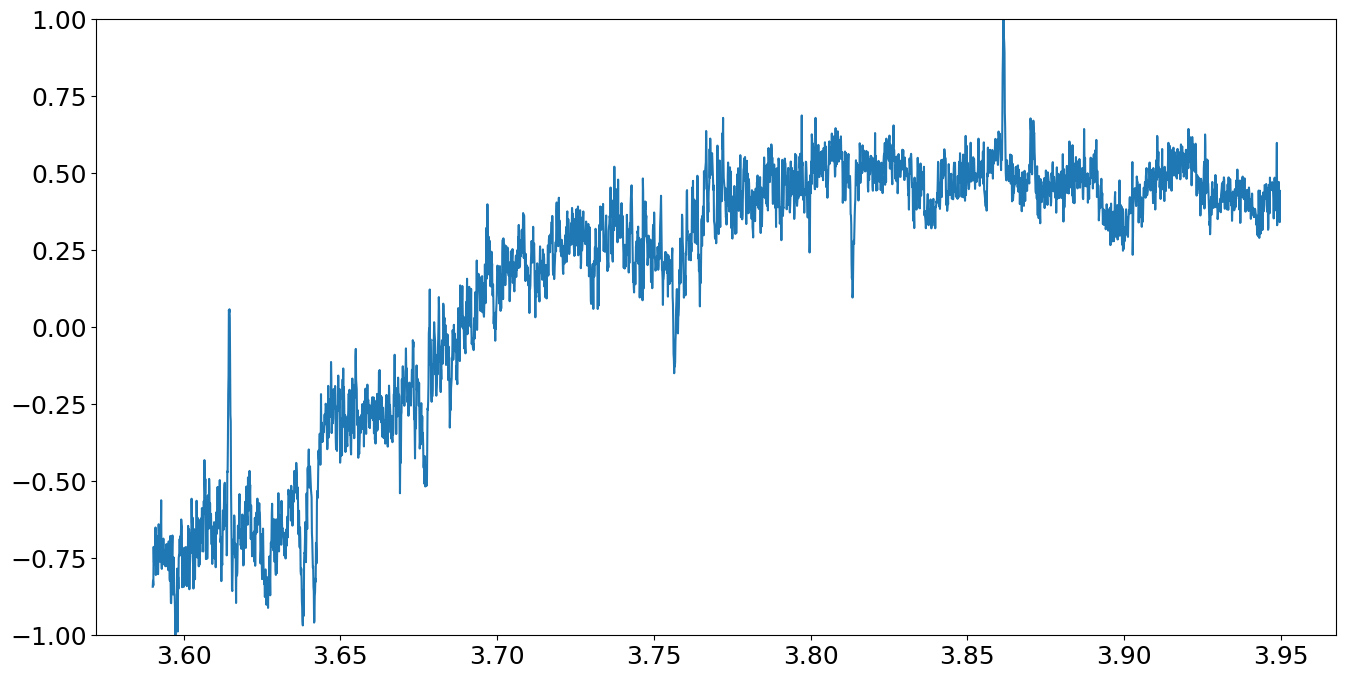} &
    \includegraphics[width=0.23\linewidth]{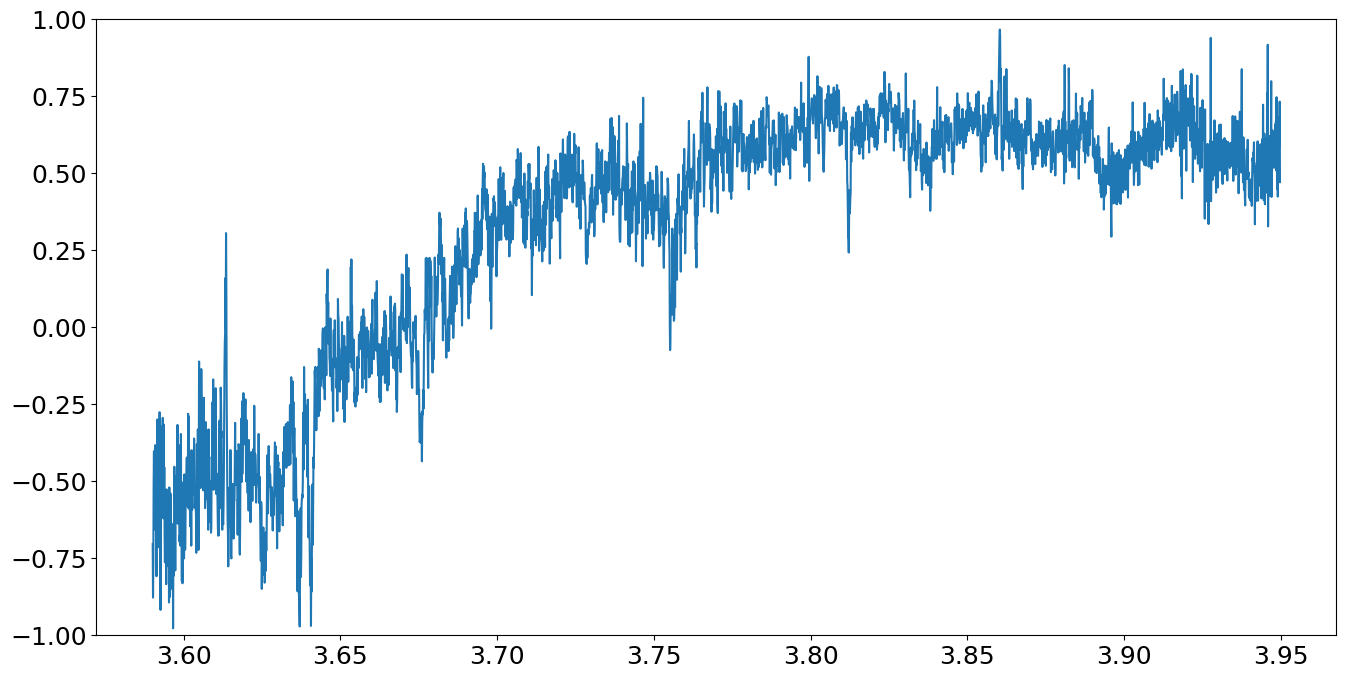} &
    \includegraphics[width=0.23\linewidth]{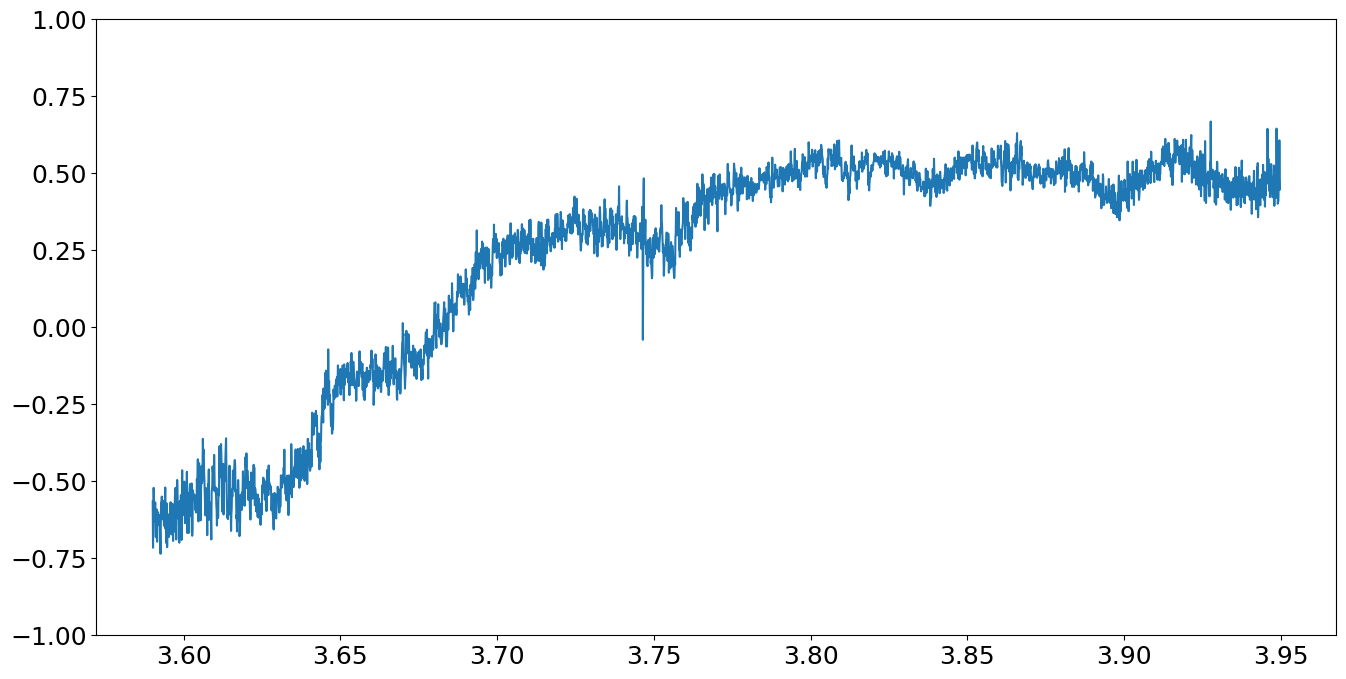}  &
    \includegraphics[width=0.23\linewidth]{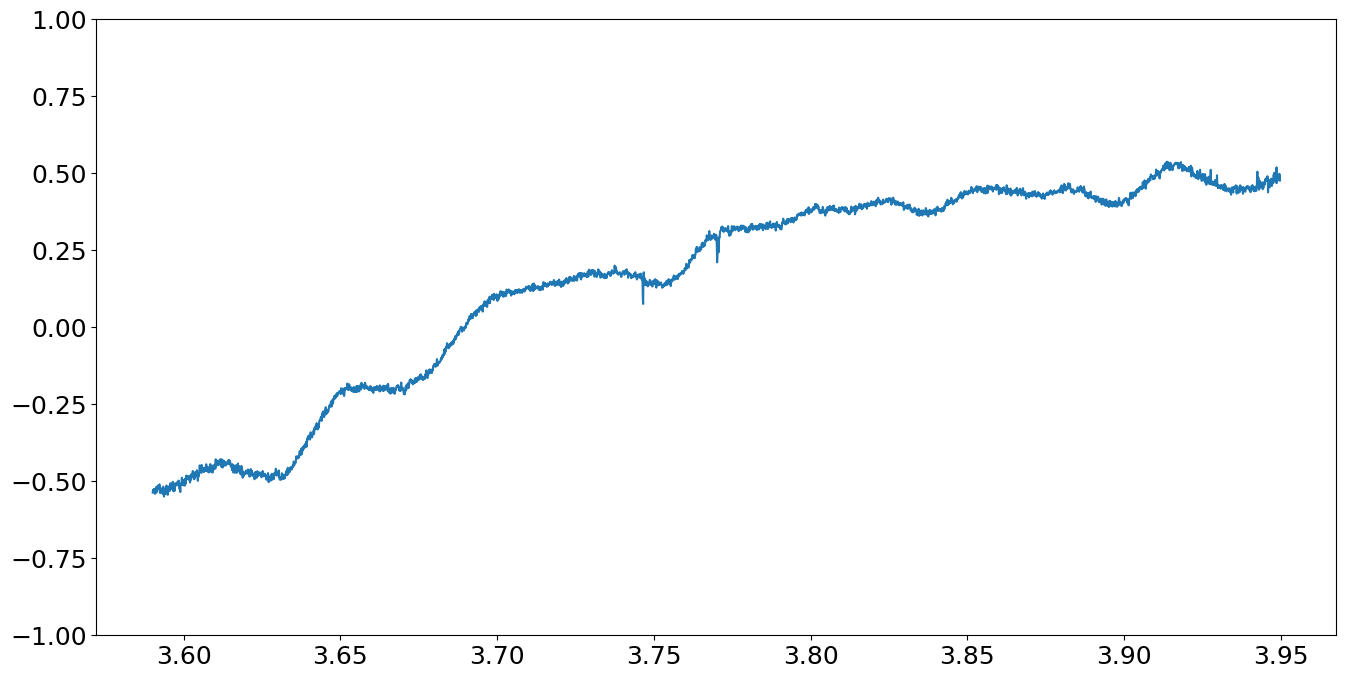}\\
    \includegraphics[width=0.08\linewidth]{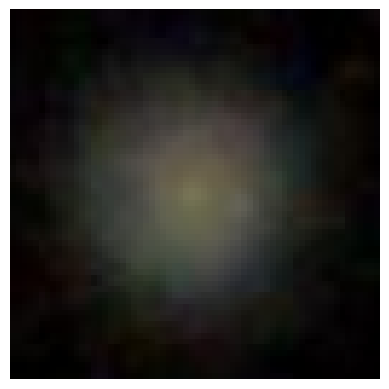} &
    \includegraphics[width=0.23\linewidth]{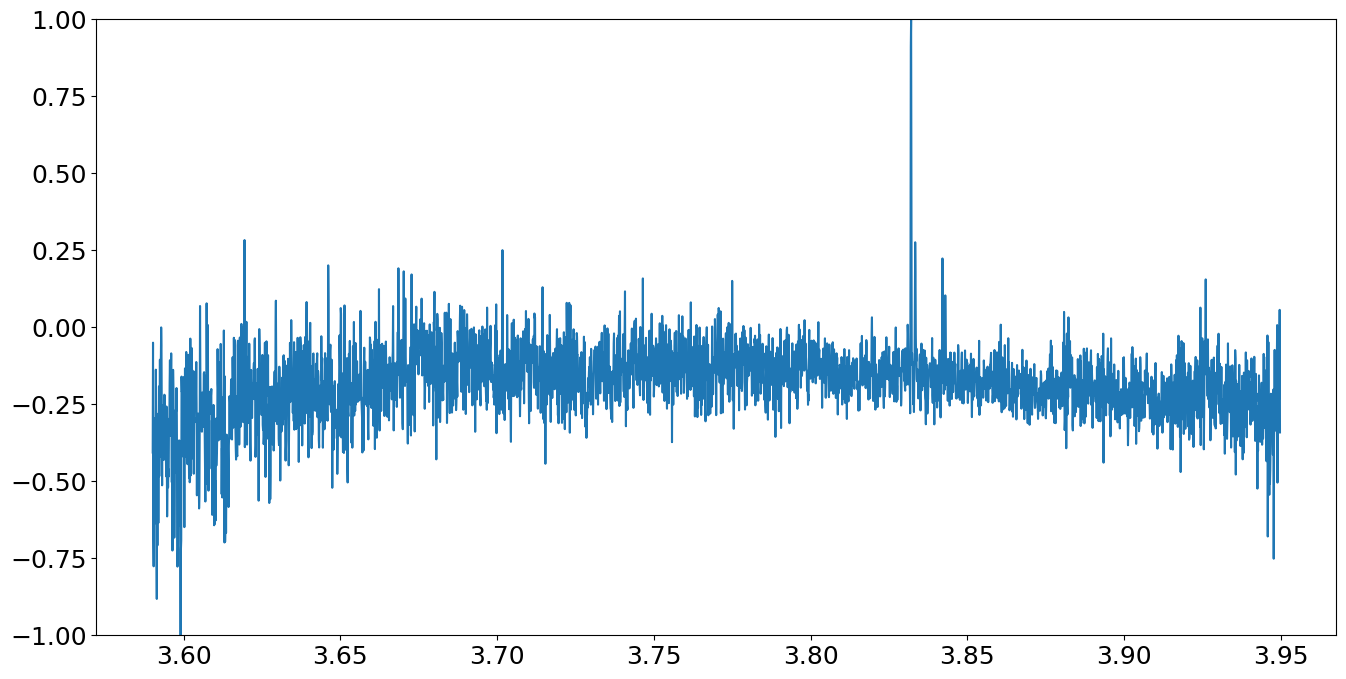} &
    \includegraphics[width=0.23\linewidth]{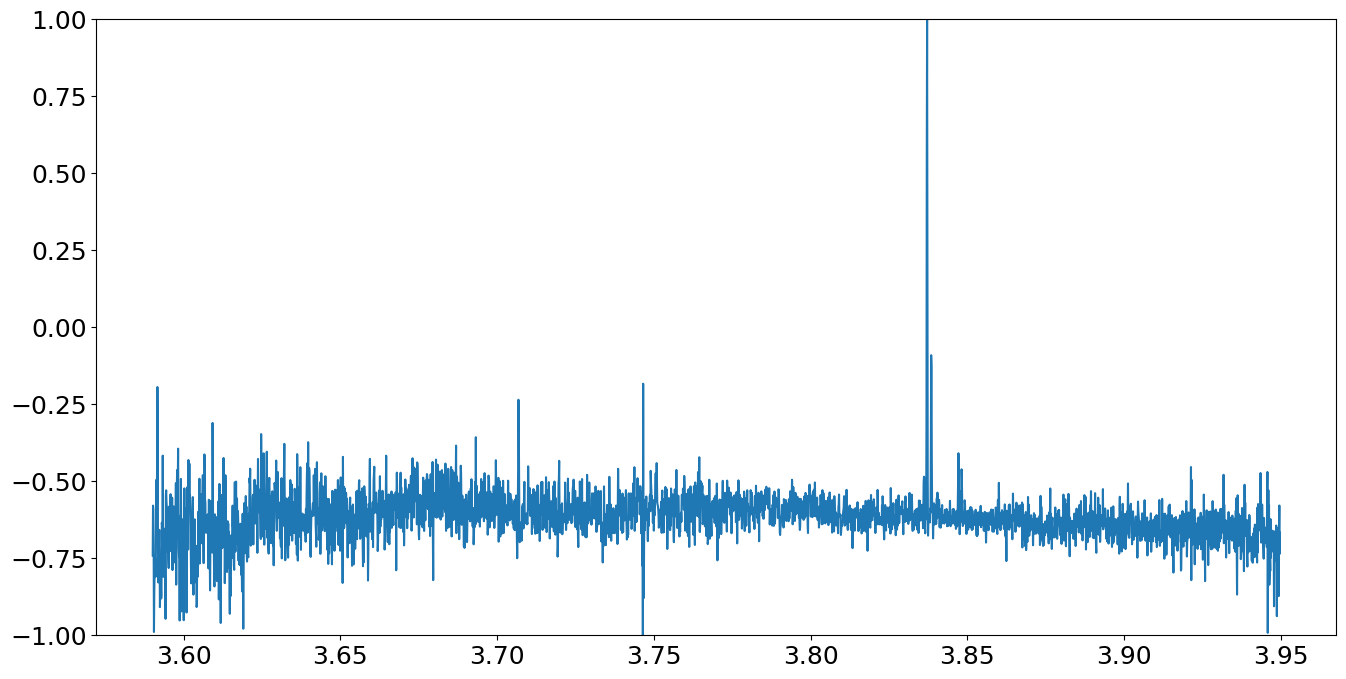} &
    \includegraphics[width=0.23\linewidth]{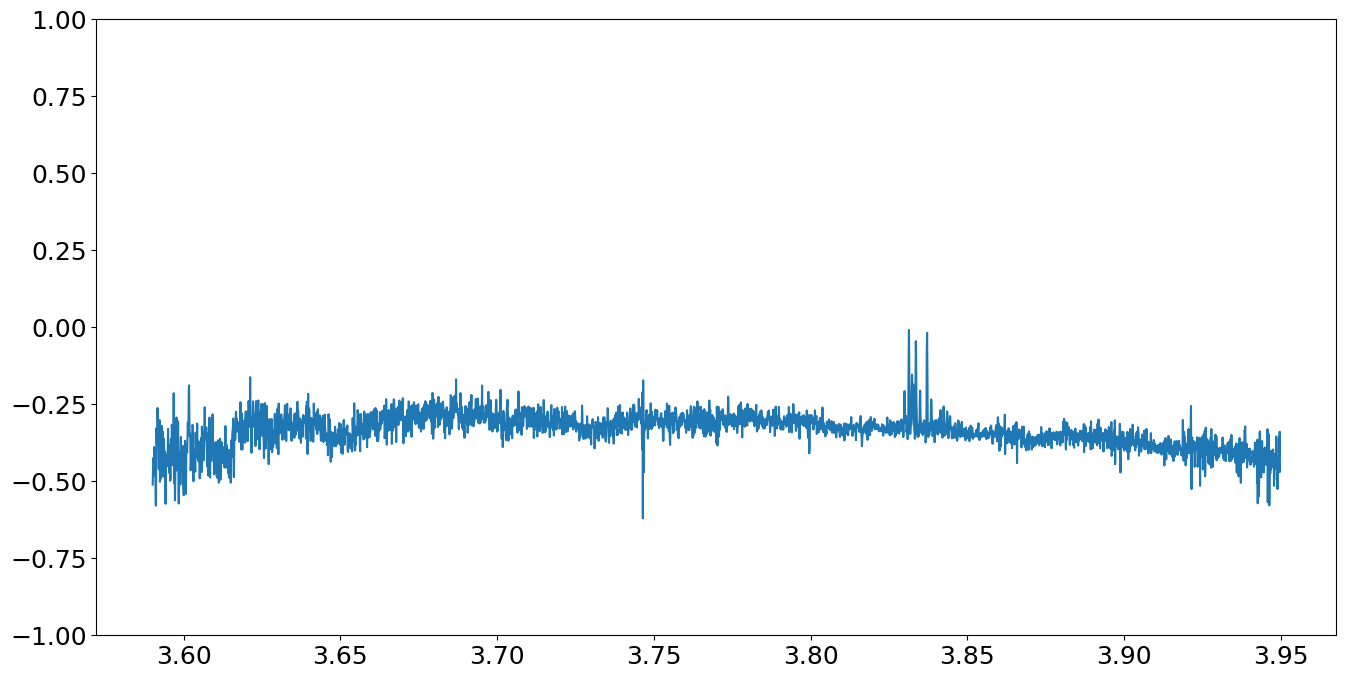} &
    \includegraphics[width=0.23\linewidth]{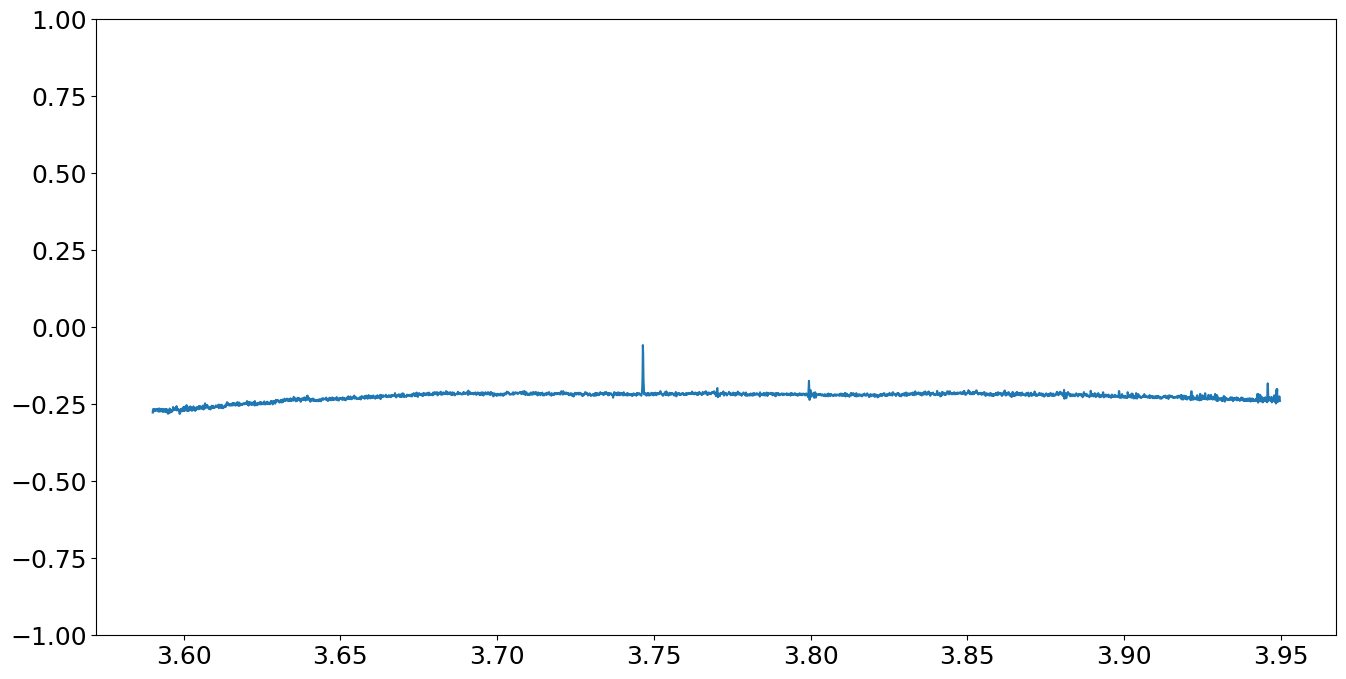}\\
    (a) & (b) & (c) & (d) & (e)\\
  \end{tabular}
    \caption{Generated spectra for the images in (a). In (b) we show the target (real) spectra, in (c) the best match according to our contrastive model out of 25 samples, in (d) the spectrum obtained by averaging the 5 best matches according to the contrastive model, out of 25 samples total, and in (e) the results with the method of \citet{wu2020predicting}. While the MSE of (e) is the lowest, we believe (c) to be the most useful.}
    \label{fig:results}
\end{figure}

\paragraph{Implementation details} We normalize the spectra to the range $[-1, 1]$. All images are resized to 71x71, random cropped to 64x64 and randomly flipped horizontally and vertically with a probability of 0.5. We train the CDM using Adam~\citet{kingma2014adam} with a learning rate of $10^{-4}$, and a batch size of 256. We use 250 timesteps with a cosine variance schedule. Models are trained for 500k iterations with 2 NVIDIA GeForce RTX 3090s. The contrastive network is trained with Adam, a learning rate of $10^{-4}$, a weight decay of $10^{-3}$, and a batch size of 512.

\paragraph{Compared methods} We compare our method against that of \citet{wu2020predicting}. They first train the VAE of \citet{portillo2020dimensionality} to learn a distribution over spectra, followed by a network that predicts the latent code of a spectrum from the associated image. Moreover, we adapt ULISSE \citet{doorenbos2022ulisse} for this use-case. ULISSE only uses photometry, and finds objects sharing similar properties by a similarity search in pretrained feature space. To obtain a guess of a spectrum, we find the nearest neighbour of an image in the feature space of a pretrained EfficientNet-b0, and use its spectrum as a guess for the spectrum of the query.

\paragraph{Results} The quantitative results in Table~\ref{tab:results} show that selecting the best spectrum with the contrastive network improves upon generating single spectra in mean squared error (MSE). We outperform ULISSE, but the method of \citet{wu2020predicting} reaches a lower MSE than ours. The qualitative results shown in Figure~\ref{fig:results} suggest this reduction in MSE is caused by it producing much smoother, less realistic samples. Metrics like the MSE are known to favor smooth samples \citet{zhang2018unreasonable}, and are thus not a perfect metric for our goal. For instance, by taking the mean of the best matching samples, we can reduce the MSE, but the samples become interpolations of multiple realistic spectra.

The fifth row in Figure~\ref{fig:results} shows an unfortunate consequence of the normalization to $[-1, 1]$. Our generated spectrum in column (c) generally lies around 0.25 lower than the actual spectrum in (b), although their shape is similar. Because normalization is done by $x_{norm} = \frac{x - x_{min}}{x_{max} - x_{min}}$, the extrema of a spectrum have a large impact on what values $x_{norm}$ will contain. Hence, the relative size of the peak around $\log \lambda \pm 3.83$ would need to be estimated perfectly to get the whole spectrum in the right place. If done incorrectly, this has a very large negative impact on MSE, despite an otherwise reasonably well-estimated spectrum.

\begin{table}
\caption{Mean and standard deviation over 3 runs of the MSE on the validation set. Contrastive ($n$) uses the contrastive network to find the best match out of $n$ samples, Contrastive+Mean ($n$/$m$) averages the best $m$ out of $n$ matches according to the contrastive network.}
\begin{center}
\hskip-0.0cm
\begin{tabular}{cc|cc|c}
\hline\hline
$\epsilon_\theta$ & $\tau_\theta$ & Conditioning & Selection & MSE\\
\hline
\multirow{4}{*}{\centering \makecell{U-net \\ \citet{falk2019u} \\ (12.4m)}} & \multirow{4}{*}{\centering \makecell{ResNet-18 \\ \citet{he2016deep} \\ (11.2m)}} & Concat & Single & 0.179\tiny{$\pm$0.017}\\
& & Concat & Contrastive (5) & 0.160\tiny{$\pm$0.010}\\
& & Concat & Contrastive (25) & 0.154\tiny{$\pm$0.004} \\
& & Concat & Contrastive+mean (25/5) & 0.105\tiny{$\pm$0.003}\\
\hline
\multirow{4}{*}{\centering \makecell{U-net \\ \citet{falk2019u} \\ (23.7m)}} & \multirow{4}{*}{\centering \makecell{ResNet-18 \\ \citet{he2016deep} \\ (11.2m)}} & X-attention & Single & 0.225\tiny{$\pm$0.014}\\
& & X-attention & Contrastive (5) & 0.136\tiny{$\pm$0.008}\\
& & X-attention & Contrastive (25) & 0.118\tiny{$\pm$0.010}\\
& & X-attention & Contrastive+Mean (25/5) & 0.103\tiny{$\pm$0.008}\\
\hline 
\multicolumn{4}{c}{ULISSE (5.3m)}   & 0.168\tiny{$\pm$0.000} \\
\multicolumn{4}{c}{\citet{wu2020predicting} (25.6m VAE + 6.8m hybrid-conv)} & 0.080\tiny{$\pm$0.000}\\
\hline\hline
\end{tabular}
\end{center}
\label{tab:results}
\end{table}

\section{Discussion \& conclusion}
\label{sec:dis}

Generating high-dimensional, noisy spectra is a difficult task, especially as we do not employ any further preprocessing, besides the normalization to $[-1, 1]$. Other works involving spectra and deep learning typically use a variety of preprocessing methods, such as low-pass median filtering or de-redshifting \cite{muthukrishna2019dash,portillo2020dimensionality}. However, all preprocessing imposes some assumptions on the data. As we aim to show our method works on as general a case as possible, we keep the preprocessing minimal. 
Nonetheless, with task-specific knowledge, the results can likely be improved further. For example, the images can at times be extremely faint, thus a selection based on \textit{e.g.} redshift would likely boost performance.

A direct consequence of our normalization is that we are effectively matching images with the correct \textit{shape} of spectrum, while ignoring the intensities. This is done for well-behaved training, as done in other deep learning works involving spectra such as \cite{tan2022robust}. However, this might destroy some important properties for certain use-cases. 

Additionally, despite a wide hyperparameter search, we found the VAE architecture of \citet{portillo2020dimensionality} unable to generate high-quality reconstructions of our spectra, producing overly smooth, non-realistic samples, unlike the diffusion models. We believe this is caused by the higher dimensional spectra (3598 vs. 1000) and less preprocessing. As \citet{wu2020predicting} rely on this VAE, their method will have this problem as well. Nonetheless, the fact that their MSE is lower shows a mismatch between the metric and our desired output. This discrepancy prevents proper evaluation of methods, and solving this is a crucial step in furthering this line of research.

All in all, our method is capable of producing reasonable estimates of spectra directly from photometric observations. With task-specific knowledge and pre-processing, we envision this to be an exceptionally useful tool for the selection of objects for spectroscopic follow-up.

\section{Broader impact}
\label{sec:impact}
We have applied diffusion models, a class of generative models, to an astronomical use-case. While generative models as a whole come with substantial ethical dilemmas, such as generating deepfakes or other harmful content, there are no such clear cases for our domain. Nonetheless, our application might lead to a better understanding of these models, and shed some light on how and where they can be used. On the brighter side, our generated spectra might allow for more targeted searches, thus reducing superfluous spectral acquisition, saving time, carbon and money. 

\begin{ack}
This work was funded by the Swiss National Science Foundation (SNSF), research grant 200021\_192285 “Image data validation for AI systems”.
\end{ack}

\bibliographystyle{plainnat}
\bibliography{main}

\appendix

\section{Checklist}

\begin{enumerate}

\item For all authors...
\begin{enumerate}
  \item Do the main claims made in the abstract and introduction accurately reflect the paper's contributions and scope?
    \answerYes{}
  \item Did you describe the limitations of your work?
    \answerYes{See the results paragraph in \ref{sec:exp} and \ref{sec:dis}, where we e.g. discuss that our normalization might be a limiting factor. }
  \item Did you discuss any potential negative societal impacts of your work?
    \answerYes{See Section~\ref{sec:impact}}
  \item Have you read the ethics review guidelines and ensured that your paper conforms to them?
    \answerYes{}
\end{enumerate}

\item If you are including theoretical results...
\begin{enumerate}
  \item Did you state the full set of assumptions of all theoretical results?
    \answerNA{}
        \item Did you include complete proofs of all theoretical results?
    \answerNA{}
\end{enumerate}

\item If you ran experiments...
\begin{enumerate}
  \item Did you include the code, data, and instructions needed to reproduce the main experimental results (either in the supplemental material or as a URL)?
    \answerYes{}
  \item Did you specify all the training details (e.g., data splits, hyperparameters, how they were chosen)?
    \answerYes{}
        \item Did you report error bars (e.g., with respect to the random seed after running experiments multiple times)?
    \answerYes{Standard deviations are given in Table~\ref{tab:results}}
        \item Did you include the total amount of compute and the type of resources used (e.g., type of GPUs, internal cluster, or cloud provider)?
    \answerYes{}
    \end{enumerate}

\item If you are using existing assets (e.g., code, data, models) or curating/releasing new assets...
\begin{enumerate}
  \item If your work uses existing assets, did you cite the creators?
    \answerYes{}
  \item Did you mention the license of the assets?
    \answerNA{}
  \item Did you include any new assets either in the supplemental material or as a URL?
    \answerYes{}
  \item Did you discuss whether and how consent was obtained from people whose data you're using/curating?
    \answerNA{}
  \item Did you discuss whether the data you are using/curating contains personally identifiable information or offensive content?
    \answerNA{}
\end{enumerate}

\item If you used crowdsourcing or conducted research with human subjects...
\begin{enumerate}
  \item Did you include the full text of instructions given to participants and screenshots, if applicable?
    \answerNA{}
  \item Did you describe any potential participant risks, with links to Institutional Review Board (IRB) approvals, if applicable?
    \answerNA{}
  \item Did you include the estimated hourly wage paid to participants and the total amount spent on participant compensation?
    \answerNA{}
\end{enumerate}

\end{enumerate}




\end{document}

%% file: defs.tex
\DeclareMathOperator{\EX}{\mathbb{E}}